\documentclass[twocolumn,prl,showpacs,superscriptaddress]{revtex4}

\usepackage{graphicx}
\usepackage{dcolumn}
\usepackage{bm}

\begin{document}
\title{Is the Anti-Pfaffian a PH-Pfaffian Topological Order State? }
\author{Jian Yang}
%\email{jyangmay1@yahoo.com}
\affiliation{Spinor Field LLC, Sugar Land, TX 77479, USA}
%\date{}

\begin{abstract}\
Recently we have shown that the anti-Pfaffian state and a state described by the second Landau level (SLL) projection of antiholomorphic Pfaffian wavefunctions have large overlap and almost identical low-energy orbital entanglement structures, suggesting they have the same topological order. In this paper, we further show that the similar entanglement structure observed in the SLL projected state is also identifiable in other Landau level projected states, indicating it is not the result of the SLL projection but rather a "fingerprint" of the PH-Pfaffian topological order of the original unprojected antiholomorphic Pfaffian wavefunctions. Consequently, we argue that the SLL projected state, and therefore the anti-Pfaffian state, is a PH-Pfaffian topological order state. We also discuss the implications on the edge physics.

\end{abstract}
 \maketitle

The leading candidates for the ground state of the $\nu=5/2$ fractional quantum Hall effect (FQHE) \cite{Willett} are the Pfaffian state\cite{MR}, the anti-Pfaffian state\cite{Levin}\cite{Lee} which is the particle-hole (PH) conjugate of the Pfaffian state, and the PH-Pfaffian state\cite{Son}\cite{Zucker}\cite{Yang}. There is a large number of numerical studies to support the Pfaffian state or the anti-Pfaffian state as a viable candidate for the ground state, however the same cannot be said for the PH-Pfaffian state. 

The main reason that the PH-Pfaffian state has attracted considerable attention, despite of its scant numerical support, is because its edge structure supports a fractional $\nu = 1/2$ downstream charged boson edge mode and a counterpropagating upstream neutral Majorana fermion edge mode, making it the only state among the three leading candidates with the edge structure that is consistent with the experiments\cite{Banerjee}\cite{Dutta}. The PH-Pfaffian state can be described by the following form of a generalized Moore-Read wave function with an antiholomorphic Pfaffian term\cite{Zucker}:
\begin{equation}
\label{PH-Pfaffian} {\Psi}_{PH-Pf} = P_{LLL}{Pf} ( \frac{1}{z_i^*-z_j^* } ) \prod\limits_{i<j}^N (z_i-z_j)^2,
\end{equation}
where $z_j = x_j+iy_j$ is the complex coordinate of the $j_{th}$ electron, $N$ is the total number of electrons, and $Pf[A]$ is the Pfaffian of an antisymmetric matrix $A$, and $P_{LLL}$ is the lowest Landau level (LLL) projection operator. According to the “bulk-edge” correspondence, the antiholomorphic Pfaffian term and the Jastrow term result in, respectively, a upstream neutral Majorana fermion edge mode and a downstream charged boson edge mode.

In the spherical geometry, the PH-Pfaffian state, which occurs at total flux $N_{\phi}= 2N-1$, is shown to have a high level of PH symmetry\cite{Yang}\cite{Balram}\cite{Mishmash}. The Pfaffian and anti-Pfaffian, which occur at $N_{\phi}= 2N-3$ and $N_{\phi}= 2N+1$ respectively, obviously lack such a symmetry. Unfortunately, the PH symmetry also turns out to be the very reason that the PH-Pfaffian state fails to represent a gapped ground state, as no consistent gapped ground state is found to exist at $N_{\phi}= 2N-1$ \cite{Yang}\cite{Balram}. In fact, the only visibly gapped states appear at $N_{\phi}= 2N-3$ and $N_{\phi}= 2N+1$\cite{Rezayi}. 

A simple way to address the PH symmetry issue is proposed recently by the author\cite{Yang1} by replacing the $P_{LLL}$ in Eq.(\ref{PH-Pfaffian}) with the second Landau level (SLL) projection operator $P_{SLL}$
\begin{equation}
\label{SLL-PH-Pfaffian} {\Psi}_{SPH-Pf} = P_{SLL}{Pf} ( \frac{1}{z_i^*-z_j^* } ) \prod\limits_{i<j}^N (z_i-z_j)^2.
\end{equation}
We call the state described by Eq.(\ref{SLL-PH-Pfaffian}) the SLL PH-Pfaffian (or SPH-Pfaffian in short) to distinguish it from the PH-Pfaffian state described by Eq.(\ref{PH-Pfaffian}). Although the SPH-Pfaffian state occurs at flux $N_{\phi}= 2N-1$, it lacks the PH symmetry because the number of orbitals in the SLL is $N_{orb} = N_{\phi} + 3 = 2N + 2$, as opposed to that of the LLL which is $N_{orb} = N_{\phi}+ 1 = 2N$. As shown in Ref.\cite{Yang1}, the SPH-Pfaffian represents a gapped, incompressible phase, and provides an excellent description for the exact ground state of finite-size SLL Coulomb-interacting electrons over a range of the short-distance interaction strength. As one can map the SPH-Pfaffian state to a LLL state, the LLL mapped SPH-Pfaffian state and the anti-Pfaffian state, both occur at flux $N_{\phi}= 2N+1$, are shown\cite{Yang1} to have large overlap and almost identical low-energy orbital entanglement structures\cite{Li}, suggesting both states have the same topological order - either the anti-Pfaffian or the PH-Pfaffian order. The question is: which one?

To answer the question and for the reasons that will become clear later, in this paper we project the antiholomorphic Pfaffian wavefunctions into other Landau levels, and study the entanglement spectrum of each of the resulting projected states. 

One can extend Eq.(\ref{PH-Pfaffian}) and Eq.(\ref{SLL-PH-Pfaffian}) to describe an arbitrary $n_{th}$ Landau level projected state by the following wavefunction
\begin{equation}
\label{nLL-PH-Pfaffian0} {\Psi}_{n} = P_{nLL}{Pf} ( \frac{1}{z_i^*-z_j^* } ) \prod\limits_{i<j}^N (z_i-z_j)^2,
\end{equation}
where $P_{nLL}$ is the $n_{th}$ Landau level projection operator ($n = 0, 1, 2, \cdot\cdot\cdot$), with ${\Psi}_{0}$ corresponding to ${\Psi}_{PH-Pf}$ in Eq.(\ref{PH-Pfaffian}) and ${\Psi}_{1}$ to ${\Psi}_{SPH-Pf}$ in Eq.(\ref{SLL-PH-Pfaffian}). However, this form of wavefunctions becomes numerically intractable with $N \geq 10$. In this paper, we will use two numerically more efficient variants of Eq.(\ref{nLL-PH-Pfaffian0}). We start with the one by multiplying a stabilization factor $\prod\limits_{i<j}^N |z_i-z_j|^2$ before applying $P_{nLL}$ \cite{Mishmash}\cite{Yang1}. In the spherical geometry, the resulting wavefunction can be written as 
\begin{equation}
\label{snPH-Pfaffian} {\Psi}_{n} = P_{nLL}{\Psi}_{Pf,b}^{\nu=-1} \prod\limits_{i<j}^N (u_iv_j-u_jv_i)^{3}, 
\end{equation}
and ${\Psi}_{Pf,b}^{\nu=-1}$ is the $\nu=-1$ bosonic Pfaffian wavefunction 
\begin{equation}
\label{sbPf} {\Psi}_{Pf,b}^{\nu=-1} = {Pf}(\frac{1}{ u_i^*v_j^*-u_j^*v_i^*}) \prod\limits_{i<j}^N (u_i^*v_j^*-u_j^*v_i^*)
\end{equation}
where $(u, v)$ are the spinor variables describing electron coordinates. 

The $n_{th}$ Landau level projection in Eq.(\ref{snPH-Pfaffian}) can be carried out by using the following two equations\cite{Wu2}
\begin{equation}
\label{SLL1} (Y_{l,m}^{q})^* = (-1)^{q+m}Y_{l,-m}^{-q},
\end{equation}
and
\begin{eqnarray}
\label{SLL2}
&& Y_{l,m}^{q}Y_{l',m'}^{q'} = \sum_{l"}(-1)^{l+l'+l"+2(q"+m")}(\frac{(2l+1)(2l'+1)}{4\pi(2l"+1)})^{\frac{1}{2}}
\nonumber \\
&&<lm,l'm'|l"m"><lq,l'q'|l"q">Y_{l",m"}^{q"},
\end{eqnarray}
where $q" = q+q'$ and $m" = m+m'$. In Eq.(\ref{SLL1}) and Eq.(\ref{SLL2}), $Y_{l,m}^q$ is the monopole harmonics wavefunction \cite{Wu1}\cite{Wu2} with the monopole strength $q$, the angular momentum $l = q, q+1, q+2, \cdot\cdot\cdot$, and the $z$ component of the angular momentum $-l \leq m \leq l$. 
We first expand the $\nu=-1$ bosonic Pfaffian wavefunction ${\Psi}_{Pf,b}^{\nu=-1}$ in terms of $(Y_{\frac{N}{2}-1,m}^{\frac{N}{2}-1})^*$, and $\prod\limits_{i<j}^N (u_iv_j-u_jv_i)^{3}$ in terms of $Y_{\frac{3N-3}{2},m'}^{\frac{3N-3}{2}}$ for each electron spinor coordinate. By use Eq.(\ref{SLL1}) and Eq.(\ref{SLL2}), the resulting product of two monopole harmonics wavefunctions $(Y_{\frac{N}{2}-1,m}^{\frac{N}{2}-1})^*Y_{\frac{3N-3}{2},m'}^{\frac{3N-3}{2}}$ can then be expanded as the infinite sum of single particle wavefunctions $Y_{l",m"}^{N-\frac{1}{2}}$ with $l" = N-\frac{1}{2}, N-\frac{1}{2}+1,N-\frac{1}{2}+2, \cdot\cdot\cdot$. The effect of the $P_{nLL}$ is simply carried out by retaining only the $n_{th}$ Landau level wavefunctions $Y_{N-\frac{1}{2}+n,m"}^{N-\frac{1}{2}}$, with $n = 0$ for the LLL, $n = 1$ for the SLL, and $n = 2$ for the third Landau level.

We work at a finite system with $N = 10$ electrons for the lowest three Landau levels projected states obtained from Eq.(\ref{snPH-Pfaffian}) with the total obitals $N_{orb} = N_{\phi}+1+2n = 20$, $22$, and $24$ for the LLL ($n = 0$), the SLL ($n = 1$), and the third Landau level ($n = 2$), respectively. The orbital entanglement spectrum first introduced in \cite{Li} has become an invaluable tool for characterizing topological orders. In Fig.~\ref{fig:Entanglement1}, we perform an orbital decomposition where subsystem $A$ containing half of the electrons with positive angular momentum $L_z^A$ and $B$ the other half with negative $L_z^B$, and plot the corresponding orbital entanglement spectrum as a function of the total angular momentum in subsystem $A$. The results are presented from the top panel to the bottom panel, respectively, for the anti-Pfaffian state ${\Psi}_{aPf}$, the SPH-Pfaffian state ${\Psi}_{SPH-Pf}$ ($n = 1$), the PH-Pfaffian state ${\Psi}_{PH-Pf}$ ($n = 0$), and the third Landau level projected state ${\Psi}_2$ ($n = 2$). 

As shown previously in Ref.\cite{Yang1} and seen from the top two panels, the anti-Pfaffian state ${\Psi}_{aPf}$ and the SPH-Pfaffian state ${\Psi}_{SPH-Pf}$ have almost identical low-energy entanglement spectrum structures (plotted in longer bars) - the multiplicities of the $4$ lowest entanglement levels at $L_z^A = 24.5, 25.5, 26.5, 27.5$ are $1, 1, 3$, and $5$. To our surprise, although shifted in $L_z^A$, the similar low-energy entanglement structure (again plotted in longer bars) is clearly identifiable for the other two Landau level projected states as seen from the bottom two panels - the same multiplicities of $1, 1, 3$, and $5$ at $L_z^A = 20.5, 21.5, 22.5, 23.5$ and at $L_z^A = 22.5, 23.5, 24.5, 25.5$, respectively, for the PH-Pfaffian state ${\Psi}_{PH-Pf}$ ($n = 0$) and the third Landau level projected state ${\Psi}_2$ ($n = 2$). 

%\onecolumngrid

%\begin{center}
\begin{figure}[tbhp]
\label{fig:Entanglement1}
%\vspace{0.2cm}
\includegraphics[width=\columnwidth]{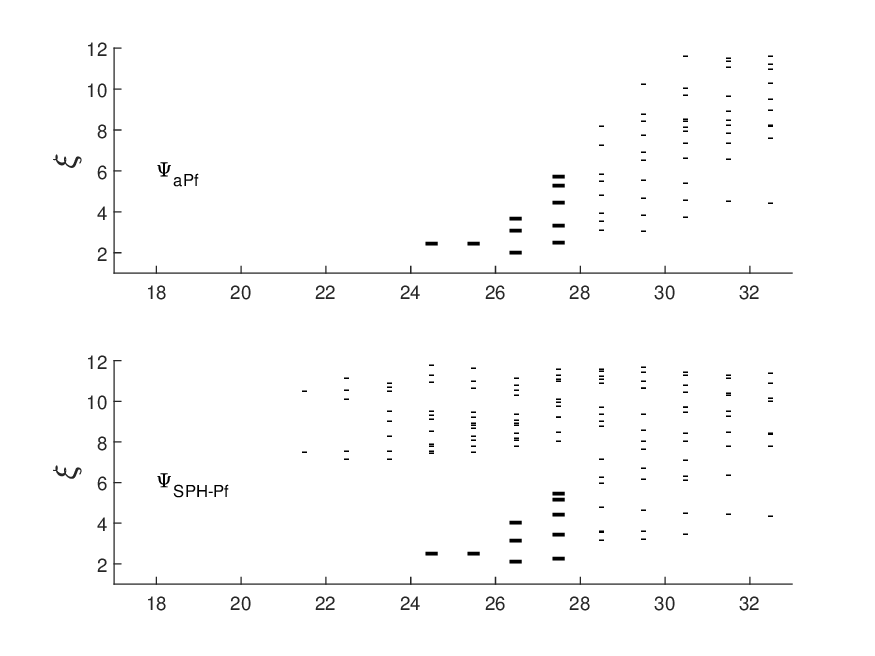}
\includegraphics[width=\columnwidth]{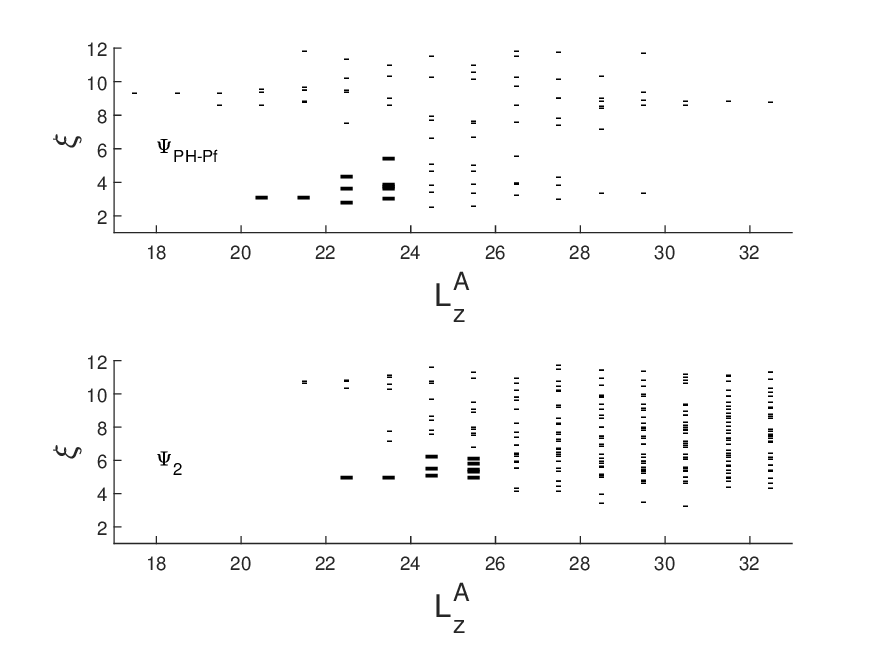}
\caption{\label{fig:Entanglement1} Orbital entanglement spectrum as a function of the total angular momentum $L_z^A$ in subsystem $A$ of a finite system with $N = 10$. The top panel is for the anti-Pfaffian state ${\Psi}_{aPf}$, and the remaining panels are for the SPH-Pfaffian state ${\Psi}_{SPH-Pf}$ ($n = 1$), the PH-Pfaffian state ${\Psi}_{PH-Pf}$ ($n = 0$), and the third Landau level projected state ${\Psi}_2$ ($n = 2$) obtained from Eq.(\ref{snPH-Pfaffian}). The low-energy entanglement spectrum with the multiplicities of $1, 1, 3$, and $5$ at the four consecutive values of $L_z^A$ are identified and plotted in longer bars.}

\end{figure}

%\end{center}
%\twocolumngrid

Next we turn to another variant of Eq.(\ref{nLL-PH-Pfaffian0}) that is numerically even more efficient than Eq.(\ref{snPH-Pfaffian}). We propose to perform the Landau level projection using the following bosonic variant of Eq.(\ref{nLL-PH-Pfaffian0}) in the spherical geometry
\begin{equation}
\label{snLL-Boson} {\Psi}_{n}^b = P_{nLL}{Pf}(\frac{1}{ u_i^*v_j^*-u_j^*v_i^*}) \prod\limits_{i<j}^N (u_iv_j-u_jv_i).
\end{equation}
To carry out the Landau level projection, we use the following equation\cite{Yutushui}
\begin{equation}
\label{SLL3} \frac{1}{ u_i^*v_j^*-u_j^*v_i^*} = \sum_{l,m} (-1)^{m-\frac{1}{2}}\frac{8\pi}{2l+1}Y_{l,m}^{\frac{1}{2}}(u_i,v_i)Y_{l,-m}^{\frac{1}{2}}(u_j,v_j),
\end{equation}
to expand ${Pf}(\frac{1}{ u_i^*v_j^*-u_j^*v_i^*})$ in terms of the single-particle wavefunctions $Y_{l,m}^{\frac{1}{2}}$ with $l = \frac{1}{2}, \frac{3}{2}, \frac{5}{2}, \cdot\cdot\cdot$, and expand $\prod\limits_{i<j}^N (u_iv_j-u_jv_i)$ in terms of the single-particle wavefunctions $Y_{l',m'}^{\frac{N-1}{2}}(u_i,v_i)$ with $l' = \frac{N-1}{2}$. As a result, the unprojected wavefunction ${Pf}(\frac{1}{ u_i^*v_j^*-u_j^*v_i^*}) \prod\limits_{i<j}^N (u_iv_j-u_jv_i)$ can be expanded in terms of $Y_{l,m}^{\frac{1}{2}}Y_{l',m'}^{\frac{N-1}{2}}$, which, by using Eq.(\ref{SLL2}) can be expressed in terms of the single particle wavefunctions $Y_{l",m"}^{\frac{N}{2}}$ with $l" = \frac{N}{2}, \frac{N}{2}+1,\frac{N}{2}+2, \cdot\cdot\cdot$. The effect of the $P_{nLL}$ is simply carried out by retaining only the $n_{th}$ Landau level wavefunctions $Y_{\frac{N}{2}+n,m"}^{\frac{N}{2}}$. One thing worth mentioning about the range of the sum over $l$ in Eq.(\ref{SLL3}). Since $l" = |l-l'|, |l-l'|+1, \cdot\cdot\cdot, l+l'$, this means $l = |l"-l'|, |l"-l'|+1, \cdot\cdot\cdot, l"+l'$. As a result, the action of $P_{nLL}$ reduces the infinite sum over $l$ in Eq.(\ref{SLL3}) to a finite sum with $l = n+\frac{1}{2}, n+\frac{3}{2}, \cdot\cdot\cdot, N+n-\frac{1}{2}$, as $l" = \frac{N}{2}+n$ and $l' = \frac{N-1}{2}$. Once the $n_{th}$ Landau level projected bosonic wavefunction ${\Psi}_{n}^b$ in Eq.(\ref{snLL-Boson}) is obtained in terms of $Y_{\frac{N}{2}+n,m"}^{\frac{N}{2}}$, the corresponding fermionic wavefunction can be obtained by mapping the projected bosonic wavefunction to the LLL, followed by multiplication with a
Vandermonde determinant $\prod\limits_{i<j}^N (u_iv_j-u_jv_i)$
\begin{equation}
\label{snLL-Fermion} {\Psi}'_{n} = (M_{LLL}{\Psi}_{n}^b) \prod\limits_{i<j}^N (u_iv_j-u_jv_i).
\end{equation}
where $M_{LLL}$ is the LLL mapping operator which can be carried out by a simple replacement from $Y_{\frac{N}{2}+n,m"}^{\frac{N}{2}}$ to $Y_{\frac{N}{2}+n,m"}^{\frac{N}{2}+n}$ for each particle coordinate. 

%\onecolumngrid

\begin{figure}[tbhp]
\label{fig:Entanglement2}
%\vspace{0.2cm}
\includegraphics[width=\columnwidth]{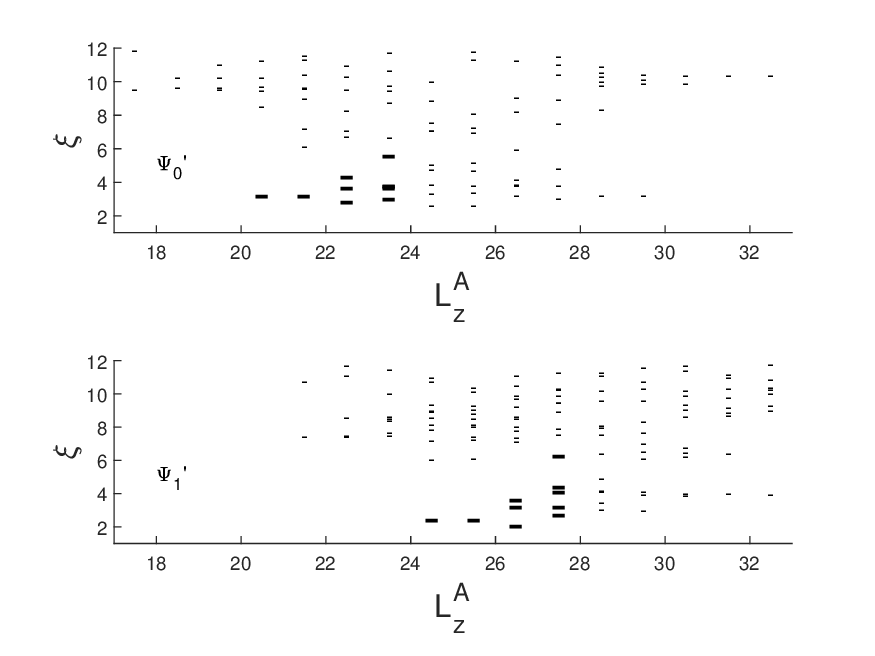}
\includegraphics[width=\columnwidth]{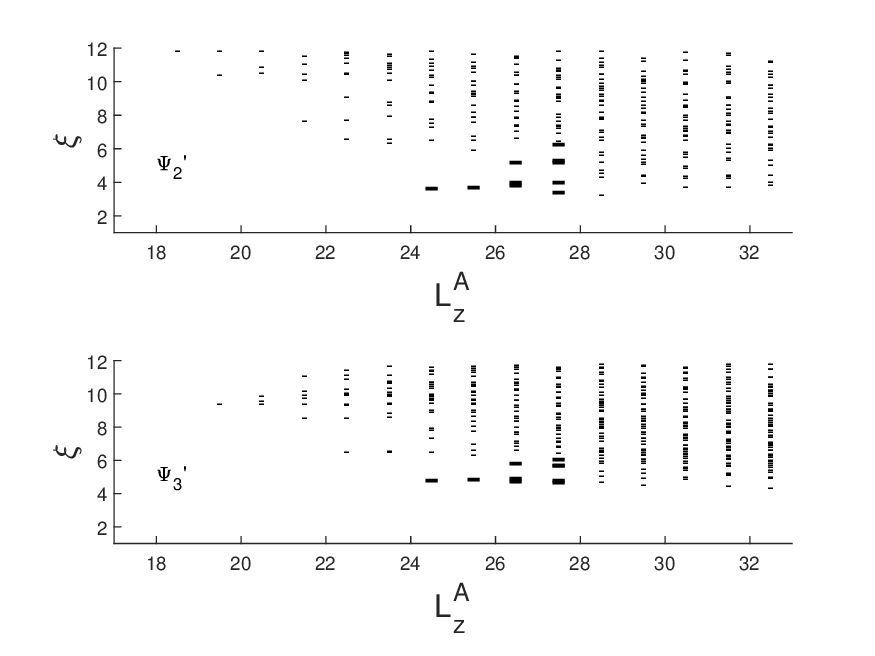}
\caption{\label{fig:Entanglement2} Orbital entanglement spectrum of a finite system with $N = 10$ electrons for the lowest four Landau level projected states ${\Psi}'_{n}$ ($n = 0, 1, 2, 3$) obtained from Eq.(\ref{snLL-Fermion}). The low-energy entanglement spectrum with the multiplicities of $1, 1, 3$, and $5$ at the four consecutive values of $L_z^A$ are identified and plotted in longer bars.}
\end{figure}

%\twocolumngrid

In Fig.~\ref{fig:Entanglement2}, we present the orbital entanglement spectrum for the lowest four Landau level projected states ${\Psi}'_{n}$ ($n = 0, 1, 2, 3$) obtained from Eq.(\ref{snLL-Fermion}). The overlaps between these states and the corresponding states obtained from Eq.(\ref{snPH-Pfaffian}) are $0.9915$, $0.9591$, and $0.3067$ for $n = 0$, $1$, and $2$ respectively. In particular the $n = 1$ state ${\Psi}'_{1}$, which can be considered as a variant of the SPH-Pfaffian state, also has a large overlap $0.9778$ with the anti-Pfaffian state. As the same as in Fig.~\ref{fig:Entanglement1}, we work at a finite system with $N = 10$ electrons, with the total orbitals $N_{orb} = N_{\phi}+1+2n = 20$, $22$, $24$, and $26$ for the LLL ($n = 0$), the SLL ($n = 1$), the third Landau level ($n = 2$), and the fourth Landau level ($n = 3$), respectively. Again, the similar low-energy entanglement structure (plotted in longer bars) are identifiable with multiplicities of $1, 1, 3$, and $5$ at $L_z^A = 20.5, 21.5, 22.5, 23.5$ for the LLL projected state ($n = 0$), and at $L_z^A = 24.5, 25.5, 26.5, 27.5$ for the other three Landau level ($n = 1$, $2$, and $3$) projected states.

Now we are ready to go back and answer the question raised in the beginning of the paper: what is the topological order of the SPH-Pfaffian state - anti-Pfaffian or the PH-Pfaffian order? We will use our numerical results to argue against one of the two possible answers and therefore in favor of the other. 

Suppose the SPH-Pfaffian state is an anti-Pfaffian topological order state, it would require that the SLL projection fundamentally alters the topological order of the original unprojected antiholomorphic Pfaffian wavefunctions, transforming it from the PH-Pfaffian order to the anti-Pfaffian order. The anti-Pfaffian entanglement structure observed in the SPH-Pfaffian state would be the result of the SLL projection, and bear no relation to the PH-Pfaffian order of the original unprojected wavefunction. For this to be true, we would expect to see different entanglement structures in different Landau level projected states. To the contrary, our numerical result presents a clear evidence that the similar entanglement structure observed in the SLL projected state is also identifiable in other Landau level projected states. 

Although in itself is not a rigorous prove, our numerical result makes the other answer much more plausible. That is, the PH-Pfaffian topological order of the original unprojected antiholomorphic Pfaffian wavefunctions survived the SLL Landau level projection, the similar entanglement structure observed in the SPH-Pfaffian state is not the result of the SLL projection, but rather a "fingerprint"\cite{Li} of the PH-Pfaffian topological order of the original unprojected antiholomorphic Pfaffian wavefunctions. Consequently, the SLL projected state is a PH-Pfaffian topological order state. Since the SPH-Pfaffian state and the anti-Pfaffian state have the same topological order, we conclude that the anti-Pfaffian state is a PH-Pfaffian topological order state, albeit without the PH symmetry. 

As a PH-Pfaffian topological order state, the anti-Pfaffian state has a PH-Pfaffian edge structure which is composed of a single edge and supports a downstream charged boson mode and an upstream neutral Majorana mode. On the other hand, the anti-Pfaffian state is also the PH conjugate of the Pfaffian state and can be regarded as a hole Pfaffian state embedded in a $\nu = 1$ integer quantum Hall state (IQHS). As a result, it also has a standard anti-Pfaffian edge structure which is composed of two edges\cite{Levin}\cite{Lee}, an edge between a hole Pfaffian bulk and $\nu = 1$ IQHS and an edge between the $\nu = 1$ IQHS and vacuum, and supports a downstream charged boson mode, an upstream neutral Majorana mode, and an upstream neutral boson edge mode. In contrast to the PH-Pfaffian edge structure, the standard anti-Pfaffian edge structure requires no knowledge about the topological order of its bulk state (other than it is the PH conjugate of the Pfaffian state), since the bulk state is shielded by the $\nu = 1$ IQHS layer without exposing to the vacuum. One may imagine as the $\nu = 1$ IQHS layer becomes narrower, the two edges will get closer and eventually merge into one edge. At some point, a transition may take place from the standard anti-Pfaffian edge structure to the PH-Pfaffian edge structure, and the upstream neutral boson edge mode that appears only in the standard anti-Pfaffian edge structure will be gapped out. The PH-Pfaffian edge structure and the standard anti-Pfaffian edge structure are mutually exclusive, while we are not in the position to determine which one of the them is more competitive energetically, so far it is the PH-Pfaffian structure that is realized in experiments\cite{Banerjee}\cite{Dutta}.

\end{document}